\documentclass[aps,prd,twocolumn,showpacs,amsmath,amssymb,nofootinbib,eqsecnum, preprintnumbers]{revtex4-1}
\usepackage{graphicx}
\usepackage{epstopdf}
\usepackage{epsfig}
\def\appendix{{\newpage\section*{Appendix}}\let\appendix\section%
        {\setcounter{section}{0}
        \gdef\thesection{\Alph{section}}}\section}
\newcommand{\be}{\begin{equation}}
\newcommand{\ee}{\end{equation}}
\newcommand{\bear}{\begin{eqnarray}}
\newcommand{\eear}{\end{eqnarray}}
\newcommand{\ba}{\begin{array}}
\newcommand{\ea}{\end{array}}

\begin{document}

\title{Configurational entropy of charged AdS black holes}
\author{Chong Oh Lee}
\email{cohlee@gmail.com}
\affiliation{Department of Physics,
Kunsan National University, Kunsan 573-701, Korea}

\begin{abstract}
When we consider charged AdS black holes in higher dimensional
spacetime and a molecule number density along coexistence curves is
numerically extended to higher dimensional cases. It is found that a
number density difference of a small and large black holes decrease
as a total dimension grows up. In particular, we find that a
configurational entropy is a concave function of a reduced
temperature and reaches a maximum value at a critical (second-order
phase transition) point. Furthermore, the bigger a total dimension
becomes, the more concave function in a configurational entropy
while the more convex function in a reduced pressure.

\end{abstract}
\maketitle
\section{Introduction}
It has been found that there is a first-order phase transition in
the Schwarzschild-AdS black hole from analogy between black hole and
standard thermodynamic system~\cite{Hawking:1982dh}. It has been
suggested that the critical behavior of charged AdS black holes
remind Van der Waals liquid-gas phase
transition~\cite{Chamblin:1999tk,Chamblin:1999hg}. Thermodynamics of
black holes in the extended space has been recently investigated by
treating the cosmological constant as the thermodynamic
pressure~\cite{Kastor:2009wy}. It has been found that the
thermodynamic pressure of charged AdS black holes is proportional to
their thermodynamic volume~\cite{Cvetic:2010jb}. The $P-V$
criticality of charged AdS black holes in the extended space has
been investigated \cite{Kubiznak:2012wp,Gunasekaran:2012dq}

A present-day concept of configurational entropy has been suggested
in Ref.~\cite{Shannon:1948zz} in search for the informational
entropy in the context of communication theory. It was recently
obtained in Ref.~\cite{Gleiser:2011di} through investigation of
measure of ordering in field configuration space for spatially
localized energy solutions of nonlinear models and used to study
instability of a variety of objects
\cite{Gleiser:2012tu,Gleiser:2013mga,Gleiser:2014ipa,
Gleiser:2015rwa,Bernardini:2016hvx,Bernardini:2016qit,Casadio:2016aum,Braga:2016wzx}.

The number density of black hole molecules was recently introduced
in Ref.~\cite{Wei:2015iwa} for investigation of measure of the
microscopic degrees of freedom of black holes. It was extensively
studied for $f(R)$ AdS black holes and Gauss-Bonnet AdS black
holes~\cite{Mo:2016sel} and for the generalization of charged AdS
black hole specific volume and number density~\cite{Wang:2017bcx}.

The paper is organized as follows: in the next section we
investigate configurational entropy in charged AdS black holes in
higher dimensional spacetime. In the four-dimensional case,
we explicitly present the configurational entropy.
In the five-dimensional/six-dimensional case, its numerical result is given.
In the last subsection we discuss their thermodynamic properties.
In the last section we give our conclusion.

\section{Configurational entropy}
In statistical thermodynamics the most general formula between the entropy and
the set of probabilities of their microscopic states
is given as the Boltzmann-Gibbs entropy $S_{BG}$
\bear\label{SBG}\textstyle
S_{BG}=-k_B \sum p_i \ln p_i,
\eear
with $\sum p_i=1$
where $k_B$ is the Boltzmann constant, and $p_i$ is the probability of a microstate.
Each microstate has equal probability as the following
\bear\textstyle
p_i=\frac{1}{W},
\eear
where $W$ is the number of microstates.
Then the Boltzmann-Gibbs entropy $S_{BG}$ (\ref{SBG}) reduces to
\bear\label{BS}\textstyle
S_{BG}=k_B \ln W,
\eear
where $W$ is treated as the number of possible configurations at the given energy,
and the Boltzmann-Gibbs entropy $S_{BG}$ (\ref{BS})
becomes the configurational entropy in the microcanonical ensemble.
Especially, supposing there are two different molecules with the total number of molecules $N_0$,
then the number of one type of molecule is $N_1$ and the number of another type of molecule
$N_1$.
The configurational entropy $S$ is written as
\bear
S=k_B \ln W=k_B \ln\left(\frac{N_0!}{N_1!\,N_2!}\right),
\eear
which leads to
\bear
S=k_B(N_0\ln N_0-N_1\ln N_1-N_2\ln N_2),
\eear
by employing sterling's approximation $\ln N!\approx N\ln N$.
Since the effective number density $n$ for the AdS black hole is given as
\cite{Kubiznak:2012wp,Gunasekaran:2012dq}
\bear
n=\frac{N}{V}=\frac{1}{2l_P^2r_h}
\eear
where $V$ is the thermodynamic volume and $l_P$ is Planck length
\bear
l_P=\sqrt{\hbar G/c^3},
\eear
the configurational entropy of charged AdS black holes per unit volume $s_{conf}$ reduces to
\bear\label{RCS}
s_{conf}=-\left[n_1\ln\left(\frac{n_1}{n_1+n_2}\right)+n_2\ln\left(\frac{n_2}{n_1+n_2}\right)\right]
\eear
by using  geometric units $G=c=k_B=\hbar=1$.

The charged AdS black hole metric in higher dimensional spacetime is given as
\bear\label{CG}\textstyle
ds^2=-f(r)dt^2+\frac{dr^2}{dr^2}+r^2 d\Omega_{d-2}^2.
\eear
with
\bear\textstyle
f(r)=1-\frac{m}{r^{d-3}}+\frac{q^2}{r^{2(d-3)}}+\frac{r^2}{l^2},
\eear
where parameter $m$ relates to the ADM mass $M$, which is identified with enthalpy
$H$ $(M\equiv H=U+PV)$~\cite{Kastor:2009wy}
\bear\textstyle
M\equiv H=\frac{\pi^{\frac{d-1}{2}}(d-2)m}{8\pi\Gamma(\frac{d-1}{2})},
\eear
and parameter $q$ relates to the black hole charge $Q$~\cite{Chamblin:1999tk,Chamblin:1999hg}
\bear\textstyle
Q=\frac{\pi^{\frac{d-1}{2}}\sqrt{2(d-2)(d-3)}q}{4\pi\Gamma(\frac{d-1}{2})}.
\eear
One may treat the cosmological constant $\Lambda$ as the thermodynamic pressure $P$
\bear\textstyle
P=-\frac{\Lambda}{8\pi}=\frac{(d-1)(d-2)}{8\pi l^2},
\eear
and its conjugate quantity as the thermodynamic volume~\cite{Cvetic:2010jb}.
\bear\textstyle
V=\frac{2\pi^{\frac{d-1}{2}}r_h^{d-1}}{\Gamma(\frac{d-1}{2})}
\eear
The $d$-dimensional black hole temperature $T$ can read
\bear\textstyle
T=\frac{1}{4\pi r_h}\bigg[(d-3)+\frac{16\pi P}{d-2}r_h^2-\frac{(d-3)q^2}{r_h^{2(d-3)}}\bigg],
\eear
which leads to the thermodynamic pressure $P$
\bear\textstyle
P=\frac{(d-2)T}{4r_h}-\frac{(d-2)(d-3)}{16\pi r_h^2}+\frac{(d-2)(d-3)q^2}{16\pi r_h^{2(d-2)}}.
\eear
Employing the Legendre transform of enthalpy $\tilde{G}=H-TS$,
the Gibbs free energy $\tilde{G}$ is given as
\bear\label{GF}\textstyle
\tilde{G}=\frac{\pi^{\frac{d-1}{2}}}{8\pi\Gamma(\frac{d-1}{2})}\bigg[r_h^{d-3}
-\frac{16\pi P r_h^{d-1}}{(d-1)(d-2)}+\frac{(2d-5)q^2}{r_h^{d-3}}\bigg].
\eear

The specific volume $v$ of the black hole fluid is identified with the horizon radius of the black hole
through comparing with the Van der Waals equation~\cite{Kubiznak:2012wp,Gunasekaran:2012dq}
\bear\textstyle
v=\frac{4l_{P}^{d-2}}{d-2},
\eear
and the equation of state is given as
\bear\label{ES1}\textstyle
P=\frac{T}{v}-\frac{d-3}{\pi(d-2)v^2}+\frac{4^{2d-5}(d-3)q^2}{4\pi(d-2)^{2d-5}v^{2(d-2)}}.
\eear

The critical point is obtained by solving the following two equations
\bear
\frac{\partial P}{\partial v}=0,~~~~~~\frac{\partial^2 P}{\partial v^2}=0,
\eear
which leads to
\bear\textstyle
P_c=\frac{(d-3)^2}{\pi(d-2)^2v_c^2},
\eear
\bear\textstyle
T_c=\frac{4(d-3)^2}{\pi(d-2)(2d-5)v_c},
\eear
\bear\textstyle
v_c=\frac{4}{d-2}\bigg[(d-2)(2d-5)q^2\bigg]^{\frac{1}{2(d-3)}},
\eear
\bear\textstyle
\tilde{G}_c=\frac{\pi^{\frac{d-1}{2}}\sqrt{(d-2)(2d-5)}q}{2\pi\Gamma(\frac{d-1}{2})(d-1)}.
\eear
Employing the the reduced physical parameters as
\bear
p=\frac{P}{P_c},~~~~~
\tau=\frac{T}{T_c},~~~~~
\nu=\frac{v}{v_c},~~~~~
G=\frac{\tilde{G}}{\tilde{G}_c},
\eear
the Gibbs free energy (\ref{GF}) is written as
\bear\label{rG}\textstyle
G=\frac{1}{4}\left[(d-1)\nu^{d-3}-\frac{(d-3)^2p\nu^{d-1}}{d-2}+\frac{d-1}{(d-2)\nu^{d-3}}\right],
\eear
and the equation of state is
\bear\textstyle
p=\frac{4(d-2)\tau}{(2d-5)\nu}-\frac{d-2}{(d-3)\nu^2}+\frac{1}{(d-3)(2d-5)\nu^{2(d-2)}},
\eear
which leads to
\bear\label{rT}\textstyle
\tau=\frac{(2d-5)p\nu}{4(d-2)}-\frac{2d-5}{4(d-3)\nu}-\frac{1}{4(d-2)(d-3)\nu^{2d-5}}.
\eear

Since the first-order phase transition occurs between the small and large black hole
along the coexistence curve except the critical point $\tau=\tau_c$, the two states
have the same  Gibbs free energy, and
Eqs. (\ref{rG}) and (\ref{rT}) are written as
\bear\label{rGS}\textstyle
G_1&=&\frac{1}{4}\left[(d-1)\nu_1^{d-3}
-\frac{(d-3)^2p\nu_1^{d-1}}{d-2}+\frac{d-1}{(d-2)\nu_1^{d-3}}\right],\nonumber\\
&=&\frac{1}{4}\left[(d-1)\nu_2^{d-3}
-\frac{(d-3)^2p\nu_2^{d-1}}{d-2}+\frac{d-1}{(d-2)\nu_2^{d-3}}\right]\nonumber\\
&=&G_2,
\eear
\bear\label{rTS}\textstyle
\tau&=&\frac{(2d-5)p\nu_1}{4(d-2)}
-\frac{2d-5}{4(d-3)\nu_1}-\frac{1}{4(d-2)(d-3)\nu_1^{2d-5}},\nonumber\\
    &=&\frac{(2d-5)p\nu_2}{4(d-2)}-\frac{2d-5}{4(d-3)\nu_2}
    -\frac{1}{4(d-2)(d-3)\nu_2^{2d-5}}.\nonumber\\
\eear

\subsection{The four-dimensional case}
In the case of $d=4$, the above Eqs. (\ref{rGS}) and (\ref{rTS}) reduce to~\cite{Wei:2015iwa,Mo:2016sel}
\bear\label{rGS4}
\frac{p\nu_1^4-6\nu_1^2-3}{8\nu_1}=\frac{p\nu_2^4-6\nu_2^2-3}{8\nu_2},
\eear
\bear\label{rTS41}
\frac{3p\nu_1^4+6\nu_1^2-1}{8\nu_1^3}=\frac{3p\nu_2^4+6\nu_2^2-1}{8\nu_2^3},
\eear
\bear\label{rTS42}
2\tau=\frac{3p\nu_1^4+6\nu_1^2-1}{8\nu_1^3}+\frac{3p\nu_2^4+6\nu_2^2-1}{8\nu_2^3}.
\eear
For convenience, we now employ the parameters as the following
\bear
x=\nu_1+\nu_2,~~~~~~y=\nu_1\nu_2,
\eear
and the above Eqs. (\ref{rGS4})$\sim$(\ref{rTS42}) are written as
\bear\textstyle
-px^2y+py^2+6y-3=0,
\eear
\bear\textstyle
2py^3+x^2-6y^2-y=0,
\eear
\bear\textstyle
-3pxy^3+x^3-6xy^2-3xy+16\tau y^3=0
\eear
These equations can be analytically solved, and the corresponding reduced pressure $p$ is obtained as
\bear
p=\frac{2^{\frac{4}{3}}\tau^2(-\tau+\sqrt{\tau^2-2})^{\frac{2}{3}}}
{[2^{\frac{1}{3}}+(-\tau+\sqrt{\tau^2-2})^{\frac{2}{3}}]^2},
\eear
which is shown as red solid curve in Fig. 1.

Introducing the number density of black hole molecules $n=1/v$, we have
\bear
\frac{n_1-n_2}{n_c}&=&\frac{\nu_2-\nu_1}{\nu_1\nu_2}=\frac{x^2-4y}{y}=\sqrt{6-6\sqrt{p}}\nonumber\\
&=&\sqrt{6-\frac{6\times2^{\frac{2}{3}}
\tau(-\tau+\sqrt{\tau^2-2})^{\frac{1}{3}}}{2^{\frac{1}{3}}+(-\tau+\sqrt{\tau^2-2})^{\frac{2}{3}}}}.
\eear
which is shown as red solid curve in Fig. 2.
Here, $n_1$ and $n_2$ are explicitly calculated as
\bear\textstyle
n_1&=&\frac{1}{v_1}=\frac{\sqrt{x^2-4y}+x}{2y}=\frac{\bigg(\sqrt{3-\sqrt{p}}
+\sqrt{3-3\sqrt{p}}\bigg)\sqrt{p}}{\sqrt{2}},\nonumber\\
&=&\frac{1}{\sqrt{2}}\left(\sqrt{3-\frac{2^{\frac{2}{3}}
\tau(-\tau+\sqrt{\tau^2-2})^{\frac{1}{3}}}{2^{\frac{1}{3}}
+(-\tau+\sqrt{\tau^2-2})^{\frac{2}{3}}}}\right.\nonumber\\
&&~~~~~~+\left.\sqrt{3-\frac{3\times2^{\frac{2}{3}}
\tau(-\tau+\sqrt{\tau^2-2})^{\frac{1}{3}}}{2^{\frac{1}{3}}
+(-\tau+\sqrt{\tau^2-2})^{\frac{2}{3}}}}
\right)\nonumber\\
&&~~~~~~~~~\times\frac{2^{\frac{2}{3}}
\tau(-\tau+\sqrt{\tau^2-2})^{\frac{1}{3}}}{2^{\frac{1}{3}}+(-\tau+\sqrt{\tau^2-2})^{\frac{2}{3}}},
\eear
\bear\textstyle
n_2&=&\frac{1}{v_2}=\frac{\sqrt{x^2-4y}-x}{2y}=\frac{\bigg(\sqrt{3-\sqrt{p}}
-\sqrt{3-3\sqrt{p}}\bigg)\sqrt{p}}{\sqrt{2}},\nonumber\\
&=&\frac{1}{\sqrt{2}}\left(\sqrt{3-\frac{2^{\frac{2}{3}}
\tau(-\tau+\sqrt{\tau^2-2})^{\frac{1}{3}}}{2^{\frac{1}{3}}
+(-\tau+\sqrt{\tau^2-2})^{\frac{2}{3}}}}\right.\nonumber\\
&&~~~~~~-\left.\sqrt{3-\frac{3\times2^{\frac{2}{3}}
\tau(-\tau+\sqrt{\tau^2-2})^{\frac{1}{3}}}{2^{\frac{1}{3}}
+(-\tau+\sqrt{\tau^2-2})^{\frac{2}{3}}}}
\right)\nonumber\\
&&~~~~~~~~~\times\frac{2^{\frac{2}{3}}
\tau(-\tau+\sqrt{\tau^2-2})^{\frac{1}{3}}}{2^{\frac{1}{3}}+(-\tau+\sqrt{\tau^2-2})^{\frac{2}{3}}}.
\eear
Substituting with the configurational entropy (\ref{RCS}), we get
\bear\textstyle
s_{conf}&=&-\Bigg [\frac{\bigg(\sqrt{3-\sqrt{p}}
+\sqrt{3-3\sqrt{p}}\bigg)\sqrt{p}}{\sqrt{2}}\nonumber\\
&&~~~\times\ln\left\{\frac{1}{2}\left(1+\frac{\sqrt{3-3\sqrt{p}}}{\sqrt{3-\sqrt{p}}}\right)\right\}\nonumber\\
&&~~~~~~+\frac{\bigg(\sqrt{3-\sqrt{p}}
-\sqrt{3-3\sqrt{p}}\bigg)\sqrt{p}}{\sqrt{2}}\nonumber\\
&&~~~~~~~~~
\times\ln\left\{\frac{1}{2}\left(1-\frac{\sqrt{3-3\sqrt{p}}}{\sqrt{3-\sqrt{p}}}\right)\right\}\Bigg]\nonumber\\
&=&-\Bigg[
\frac{1}{\sqrt{2}}\left(\sqrt{3-\frac{2^{\frac{2}{3}}
\tau(-\tau+\sqrt{\tau^2-2})^{\frac{1}{3}}}{2^{\frac{1}{3}}
+(-\tau+\sqrt{\tau^2-2})^{\frac{2}{3}}}}\right.\nonumber\\
&&~~~~~~+\left.\sqrt{3-\frac{3\times2^{\frac{2}{3}}
\tau(-\tau+\sqrt{\tau^2-2})^{\frac{1}{3}}}{2^{\frac{1}{3}}
+(-\tau+\sqrt{\tau^2-2})^{\frac{2}{3}}}}
\right)\nonumber\\
&&~~~~~~~~~\times\frac{2^{\frac{2}{3}}
\tau(-\tau+\sqrt{\tau^2-2})^{\frac{1}{3}}}{2^{\frac{1}{3}}+(-\tau+\sqrt{\tau^2-2})^{\frac{2}{3}}}\nonumber\\
&&\times\ln\left\{\frac{1}{2}\left(1+\frac{\sqrt{3-3\times\frac{2^{\frac{2}{3}}
\tau(-\tau+\sqrt{\tau^2-2})^{\frac{1}{3}}}{2^{\frac{1}{3}}
+(-\tau+\sqrt{\tau^2-2})^{\frac{2}{3}}}}}{\sqrt{3-\frac{2^{\frac{2}{3}}
\tau(-\tau+\sqrt{\tau^2-2})^{\frac{1}{3}}}{2^{\frac{1}{3}}
+(-\tau+\sqrt{\tau^2-2})^{\frac{2}{3}}}}}\right)\right\}\nonumber\\
&&+\frac{1}{\sqrt{2}}\left(\sqrt{3-\frac{2^{\frac{2}{3}}
\tau(-\tau+\sqrt{\tau^2-2})^{\frac{1}{3}}}{2^{\frac{1}{3}}
+(-\tau+\sqrt{\tau^2-2})^{\frac{2}{3}}}}\right.\nonumber\\
&&~~~~~~-\left.\sqrt{3-\frac{3\times2^{\frac{2}{3}}
\tau(-\tau+\sqrt{\tau^2-2})^{\frac{1}{3}}}{2^{\frac{1}{3}}
+(-\tau+\sqrt{\tau^2-2})^{\frac{2}{3}}}}
\right)\nonumber\\
&&~~~~~~~~~\times\frac{2^{\frac{2}{3}}
\tau(-\tau+\sqrt{\tau^2-2})^{\frac{1}{3}}}{2^{\frac{1}{3}}+(-\tau+\sqrt{\tau^2-2})^{\frac{2}{3}}}\nonumber\\
&&\times\ln\left\{\frac{1}{2}\left(1-\frac{\sqrt{3-3\times\frac{2^{\frac{2}{3}}
\tau(-\tau+\sqrt{\tau^2-2})^{\frac{1}{3}}}{2^{\frac{1}{3}}
+(-\tau+\sqrt{\tau^2-2})^{\frac{2}{3}}}}}{\sqrt{3-\frac{2^{\frac{2}{3}}
\tau(-\tau+\sqrt{\tau^2-2})^{\frac{1}{3}}}{2^{\frac{1}{3}}
+(-\tau+\sqrt{\tau^2-2})^{\frac{2}{3}}}}}\right)\right\},\nonumber\\
\eear
which is shown as red solid curve in Fig. 3.
\begin{figure}[!htbp]
\begin{center}
{\includegraphics[width=8cm]{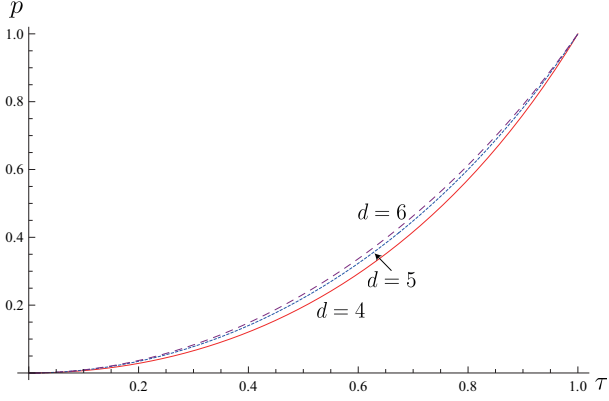}}
\end{center}
\vspace{-0.6cm}
\caption{{\footnotesize Plot of the reduced pressure $p$ as the function of the reduced temperature $\tau$
(red solid curve for $d=4$,
blue dotted curve for $d=5$, and purple dashed curve for $d=6$, respectively).}}
\label{figI}
\end{figure}
\subsection{The five-dimensional case}
As discussed in the previous section we will apply a similar analysis to the five-dimensional case.

\bear
\frac{p\nu_1^6-3\nu_1^4-1}{3\nu_1^2}=\frac{p\nu_2^6-3\nu_2^4-1}{3\nu_2^2},
\eear
\bear
\frac{10p\nu_1^6+15\nu_1^4-1}{24\nu_1^5}=\frac{10p\nu_2^6+15\nu_2^4-1}{24\nu_2^5},
\eear
\bear
2\tau=\frac{10p\nu_1^6+15\nu_1^4-1}{24\nu_1^5}+\frac{10p\nu_2^6+15\nu_2^4-1}{24\nu_2^5},
\eear
which leads to
\bear\textstyle
-px^2y^2+2py^3+3y^2-1=0
\eear
\bear\textstyle
10py^5+x^4-3x^2y-15y^4+y^2=0
\eear
\bear\textstyle
10pxy^5-x^5+5x^3y+15xy^4-5xy^2+48\tau y^5=0,\nonumber\\
\eear
which is a complicated high-order polynomial equation
and it is difficult to analyze exactly.
However some numerical investigation can be employed
and the reduced pressure $p$ as the function of the reduced temperature $\tau$ is
numerically calculated as blue dotted curve in Fig. 1.
The number density difference of the small, and large black holes $(n_1-n_2)/n_c$
as the function of the reduced temperature $\tau$ is
numerically obtained
as blue dotted curve in Fig. 2, and the configurational entropy $s_{conf}$
as the function of the reduced temperature $\tau$ is numerically given
as blue dotted curve in Fig. 3.
\begin{figure}[!htbp]
\begin{center}
{\includegraphics[width=8cm]{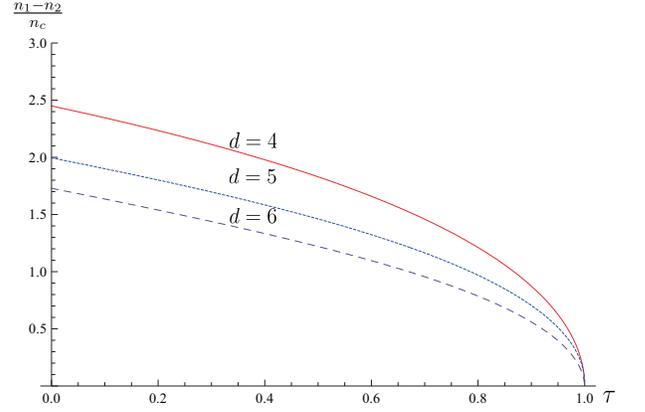}}
\end{center}
\vspace{-0.6cm}
\caption{{\footnotesize Plot of the number density difference of the small and large black holes
$(n_1-n_2)/n_c$ as the function of the reduced temperature $\tau$
(red solid curve for $d=4$,
blue dotted curve for $d=5$, and purple dashed curve for $d=6$, respectively).}}
\label{figII}
\end{figure}

\begin{figure}[!htbp]
\begin{center}
{\includegraphics[width=8cm]{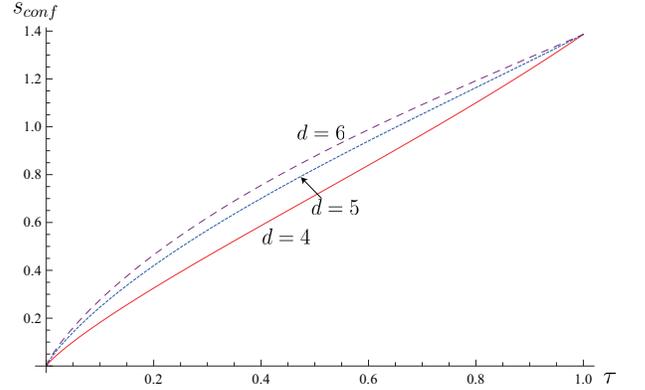}}
\end{center}
\vspace{-0.6cm}
\caption{{\footnotesize Plot of the reduced configurational entropy $s_{conf}$
as the function of the reduced temperature $\tau$
(red solid curve for $d=4$,
blue dotted curve for $d=5$, and purple dashed curve for $d=6$, respectively).}}
\label{figIII}
\end{figure}
\subsection{The six-dimensional case}
We now apply similar numerical investigations to the six-dimensional case.
\bear
\frac{9p\nu_1^8-20\nu_1^6-5}{16\nu_1^3}=\frac{9p\nu_2^8-20\nu_2^6-5}{16\nu_2^3},
\eear
\bear
\frac{21p\nu_1^8+28\nu_1^6-1}{48\nu_1^7}=\frac{21p\nu_2^8+28\nu_2^6-1}{48\nu_2^7},
\eear
\bear
2\tau=\frac{21p\nu_1^8+28\nu_1^6-1}{48\nu_1^7}+\frac{21p\nu_2^8+28\nu_2^6-1}{48\nu_2^7},
\eear
which leads to
\bear\textstyle
&&-9px^4y^3+27px^2y^4-9py^5+20x^2y^3\nonumber\\
&&-5x^2-20y^4+5y=0,
\eear
\bear\textstyle
21py^7+x^6-5x^4y+6x^2y^2-28y^6-y^3=0,
\eear
\bear\textstyle
&&-21pxy^7+x^7-7x^5y+14x^3y^2\nonumber\\
&&-28xy^6-7xy^3+96\tau y^7=0,
\eear
which is numerically solved and the reduced pressure $p$ as the function of the reduced temperature $\tau$,
the number density difference of the small, and large black holes $(n_1-n_2)/n_c$
as the function of the reduced temperature $\tau$, and
the configurational entropy $s_{conf}$ as the function of the reduced temperature $\tau$ are given as
purple dashed curve in Fig. 1, Fig. 2, and Fig. 3, respectively.

\subsection{The thermodynamic properties}
\begin{figure}[!htbp]
\begin{center}
{\includegraphics[width=8cm]{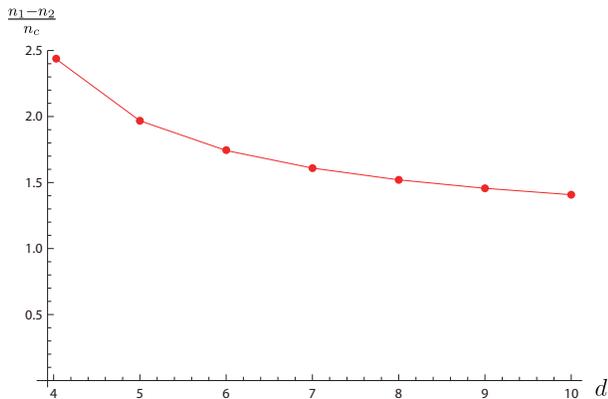}}
\end{center}
\vspace{-0.6cm}
\caption{{\footnotesize  Plot of the maximum number density difference of the small and large black holes
$(n_1-n_2)/n_c$ as the function of the total dimension $d$.}}
\label{figIV}
\end{figure}

We now discuss the thermodynamic properties in charged AdS black
holes in higher dimensional spacetime.

As shown in Fig. 1, and Fig. 2, as the reduced temperature $\tau$
grows up the reduced pressure $p$ monotonically increases while the
number density difference of the small, and large black holes
$(n_1-n_2)/n_c$ decreases. As shown in Fig. 3, the bigger the total
dimension $d$ becomes, the more concave function in the
configurational entropy $s_{conf}$. Especially, as the reduced
temperature $\tau$ grows up, $s_{cong}$ monotonically increases, and
reaches the maximum value at the critical (second-order phase
transition) point. The maximum number density difference of the
small and large black holes $(n_1-n_2)/n_c$ at $\tau=0$ is obtained
as \bear\textstyle
\frac{n_1-n_2}{n_c}=\bigg[(d-2)(2d-5)\bigg]^{\frac{1}{2(d-3)}}.
\eear As shown in Fig. 4, it decreases as the total dimension $d$
increases.

\section{Conclusion}
We considered higher dimensional charged AdS black holes and
investigated the number density difference of the small and large
black holes $(n_1-n_2)/n_c$, and the reduced configurational entropy
$s_{conf}$ in the context of the molecule number density. We
explicitly obtained the general form of maximum value of
$(n_1-n_2)/n_c$ at $\tau=0$, and found that its maximum value
decreases as the total dimension $d$ increases. Especially, the
configurational entropy $s_{conf}$ monotonically increases as the
reduced temperature $\tau$ grows up. It finally reaches a maximum
value at a critical (second-order phase transition) point. This result is
natural since in any system containing two different types of
molecules, when they have the same number of molecules, the number
of microstates $W$ reaches maximum value. Furthermore, such result
is consistent with that of the Van der Waals system. It was shown
that the critical behaviour of charged AdS black holes coincides
with those of the Van der Waals system~\cite{Kubiznak:2012wp}. In
particular, when the second-order phase transition between liquid and gas
occurs at the critical point, the distinction between the liquid and
gas phases of the Van der Waals fluid is almost non-existent near
the critical point and the molecules in the liquid and gas states
are almost identical. Then, the number of microstates $W$ becomes
maximum.

%\section*{Acknowledgements}

\end{document}